\documentclass[aps,prd,reprint,superscriptaddress,nofootinbib]{revtex4-1}
\usepackage{epsfig,amsmath,amssymb,amsfonts,verbatim,enumerate}
\usepackage{graphicx,subfigure}
\usepackage[spanish,activeacute,english]{babel}
\usepackage[latin1]{inputenc}
\usepackage{latexsym}
\usepackage{framed,color}
\usepackage[colorlinks=true,dvips]{hyperref}
\usepackage[left=1.7cm,top=2cm,right=1.5cm]{geometry}
\definecolor{shadecolor}{rgb}{0.8,0.7,0.8}
\newcommand{\beeq}{\begin{equation}}
\newcommand{\eneq}{\end{equation}}
\newcommand{\bear}{\begin{eqnarray}}
\newcommand{\enar}{\end{eqnarray}}
\newcommand{\bef}{\begin{figure*}}
\newcommand{\enf}{\end{figure*}}

\newcommand*{\mathcolor}{}
\def\mathcolor#1#{\mathcoloraux{#1}}
\newcommand*{\mathcoloraux}[3]{%
  \protect\leavevmode
  \begingroup
    \color#1{#2}#3
  \endgroup
}

\newcommand{\g}{\gamma}
\def\M{\mathcal{M}}
\def\O{\mathcal{O}}
\renewcommand{\v}[1]{\mathbf{#1}}
\newcommand{\vk}{\v{k}}

\newcommand{\Atwo}{$\tilde{b}_I^{\rm NG}A_2$}
\newcommand{\rng}{$r_n^{(\gamma)}$}
\newcommand{\dAtwo}{$\Delta (\tilde{b}_I^{\rm NG}A_2)=$}
\newcommand{\dAzero}{$\Delta A_0=$}

\newcommand{\AtwoLSSTred}{$210\,$} 
\newcommand{\AtwoEuclidred}{$220\,$}
\newcommand{\AzeroLSSTred}{$43\,$} 
\newcommand{\AzeroEuclidred}{$65\,$}

\newcommand{\AtwoLSSTredxblue}{$47\,$}

\newcommand{\AtwoLSSTrblten}{$94\,$} 
\newcommand{\AtwoEuclidrblten}{$110\,$}

\newcommand{\dAtwoLSSTredxblue}{$22\%\,$} 
\newcommand{\dAtwoapprox}{$\sim 22\%\,$} 

\newcommand{\dAtwoapproxCMB}{$\sim 40\%\,$} 

\begin{document}

\setcounter{section}{0}
\setcounter{subsection}{0}

\title{Multitracing Anisotropic Non-Gaussianity with Galaxy Shapes}
\author{Nora Elisa Chisari}
\affiliation{Department of Physics, University of Oxford, Denys Wilkinson Building, Keble Road, Oxford OX1 3RH, United Kingdom}
\email{elisa.chisari@physics.ox.ac.uk}
\author{Cora Dvorkin}
\affiliation{Department of Physics, Harvard University, Cambridge, MA 02138, USA}
\author{Fabian Schmidt}
\affiliation{Max-Planck-Institute for Astrophysics, D-85748 Garching, Germany}
\author{David N. Spergel}
\affiliation{Department of Astrophysical Sciences, Princeton University, 4 Ivy Lane, Princeton, NJ 08544, USA}

\begin{abstract}
  Correlations between intrinsic galaxy shapes on large-scales arise due to the effect of the tidal field of the large-scale structure. Anisotropic primordial non-Gaussianity induces a distinct scale-dependent imprint in these tidal alignments on large scales. Motivated by the observational finding that the alignment strength of luminous red galaxies depends on how galaxy shapes are measured, we study the use of two different shape estimators as a multi-tracer probe of intrinsic alignments. We show, by means of a Fisher analysis, that this technique promises a significant improvement on anisotropic non-Gaussianity constraints over a single-tracer method. For future weak lensing surveys, the uncertainty in the anisotropic non-Gaussianity parameter, $A_2$, is forecast to be $\sigma(A_2)\approx 50$, \dAtwoapproxCMB~smaller than currently available constraints from the bispectrum of the Cosmic Microwave Background. This corresponds to an improvement of a factor of $4-5$ over the uncertainty from a single-tracer analysis.  
\end{abstract}
\maketitle

\section{Introduction}

The next generation of galaxy surveys will mine the cosmological information in the large-scale structure of the Universe with unprecedented precision in the quest to constrain the nature of ``dark energy'', the mysterious force behind the accelerated expansion of the Universe. Two of the most promising probes of the growth history of the Universe are the clustering of galaxies and their gravitational lensing by intervening matter along the line of sight. The deviations of photons from their otherwise straight path produced by lensing result in changes in the ellipticities of galaxies of order $1\%$. 

Aside from gravitational lensing, the tidal field of the large-scale structure of the Universe can also modify the shapes and orientations of galaxies. These ``intrinsic alignments'' have been clearly detected for luminous red galaxies up to $z\sim 1$ \citep{Brown02,Mandelbaum06,Hirata07,Joachimi11,Heymans13,Singh15}, with an alignment bias that depends on luminosity. Recently, Ref.~\cite{Singh16} showed that the alignment strength also changes when different regions of a galaxy are probed. Their results suggest that the outskirts of galaxies are more sensitive to the tidal field, with their isophotes twisting more efficiently in the direction of other galaxies \cite{diTullio79,Kormendy82}. Alignments of red galaxies have also been clearly identified in cosmological hydrodynamical simulations, including the radial dependence of alignment strength \citep{Tenneti15,Velliscig15,Chisari15HzAGN,Chisari16HzAGN}. 

While intrinsic alignments are widely regarded as contaminants to weak gravitational lensing \citep{Hirata04,Joachimi10,Joachimi10b,Zhang10,Kirk15,Krause15,Krause16}, recent work has started to explore them as a cosmological probe in their own right \citep{chisari/dvorkin,CDS14,Schmidt15}. The ``linear tidal alignment model'' \citep{Catelan01,Blazek11} provides a good description of the scale and redshift dependence of intrinsic alignments on scales $\gtrsim 10$ Mpc$/h$. In \cite{Schmidt15}, hereafter SCD15, we explored the potential of intrinsic alignments as a probe of inflation; in particular, through the scale-dependent bias in the statistics of intrinsic galaxy shapes that arises in the presence of anisotropic non-Gaussianity in the early universe. Constraints on this type of non-Gaussianity are inaccessible through two-point correlations of galaxy clustering, and they probe: primordial curvature perturbations generated by large-scale magnetic fields \cite{ShiraishiB1,ShiraishiB2}, the presence of higher spin (spin $2$) fields during inflation \citep{Arkani-Hamed:2015bza,Chen/Wang:1,Chen/Wang:2,Baumann/Green,Lee16}, inflationary models with a generalized bispectrum from excited Bunch-Davies vacuum \cite{Ganc12,Agullo12,Ashoorioon16},  vector fields \cite{Yokoyama08,Barnaby12,Bartolo13,Bartolo15} and solid inflation \cite{Endlich13}.

Assuming scale invariance, the squeezed-limit bispectrum of the primordial Bardeen potential perturbation $\phi$ can in general be expressed as \cite{Shiraishi13}
\begin{eqnarray}
B_\phi(\vk_1,\vk_2,\vk_3=\vk_L) &=& \sum_{\ell=0,2,...} A_\ell P_\ell(\hat\vk_L\cdot\hat\vk_S) 
\left(\frac{k_L}{k_S}\right)^\Delta \label{eq:BsqI} \\
& &\times P_\phi(k_L) P_\phi(k_S)\left[1 + \O\left(\frac{k_L^2}{k_S^2}\right)\right]\,, \nonumber
\end{eqnarray}
where $k_3 = k_L \ll k_1,\,k_2$ while $\vk_S = \vk_1 -\vk_L/2$ and $\vk_1 + \vk_2 + \vk_L = 0$ from statistical homogeneity, $P_\ell$ are the Legendre polynomials, and $A_\ell$ are dimensionless amplitudes which are allowed to be non-zero only for even $\ell$ in the squeezed limit. The coefficient $A_0$ is related to the usual local non-Gaussianity parameter $f_{\rm NL}^{\rm loc}$ via $A_0 = 4 f_{\rm NL}^{\rm loc}$.  Intrinsic alignments constrain the parameter that governs the quadrupolar dependence of the bispectrum, $A_2$. As in SCD15, we will focus on the ``local'' scaling with $\Delta=0$. Our results are easily generalizable to other values of $\Delta$, which might be of particular relevance for massive higher-spin fields, since in de Sitter space their masses are bounded from below by a unitarity condition known as the ``Higuchi bound" \citep{Lee16}.

In SCD15, we showed that constraints on anisotropic non-Gaussianity are complementary to those derived from the Cosmic Microwave Background (CMB) bispectrum \cite{PlanckNG}, but probing smaller scales. Current constrains from Planck for the anisotropic non-Gaussianity parameter are $A_2=-16\pm86$ from temperature information only, or $A_2=6\pm74$ including preliminary polarization information ($1\sigma$, Table 25 of \cite{PlanckNG15}). Ref. \cite{Raccanelli15} explored the potential constraints on $A_2$ from biased tracers in future radio surveys. In their optimistic scenario, an uncertainty of $\Delta A_2=250$ could be reached with the Square Kilometer Array if the redshift distribution of the sources can be inferred through cross-correlations with samples of known redshift. This constraint is not competitive with the CMB because the leading contribution to the scale-dependent bias of tracer counts cancels for anisotropic primordial non-Gaussianity, as mentioned above. 
 
Future galaxy surveys, such as Euclid\footnote{\url{http://sci.esa.int/euclid/}} and the {\it Large Synoptic Survey Telescope} (LSST\footnote{\url{http://www.lsst.org/lsst/}}), will gain constraining power on cosmological parameters from applying the so-called ``multi-tracer'' technique \citep{Seljak09,McDonald09}. This technique combines tracers of the same underlying density field, with different bias parameters, to reduce the impact of cosmic variance. Note that this cosmic variance cancellation only applies to scale-dependent features in the statistics of these tracers. The application of this technique to measure ``ultra-large scale'' observables, for example, general relativistic effects and non-Gaussianity of the local type, $f_{\rm NL}$, is particularly promising with future surveys \citep{yoo/etal:2012,Ferramacho14,Fonseca15,Alonso15}.

In this work, we explore how the combination of intrinsic alignments measured from different regions of a galaxy, combined in a multi-tracer technique, can enhance constraints on anisotropic non-Gaussianity. We show that error bars can be significantly smaller than when estimated from a single-tracer method. The main result of this paper is the reduction of the uncertainty in the anisotropic non-Gaussianity parameter, $A_2$, to \dAtwoapprox~of the single-tracer value when the multi-tracer technique is applied to red and blue galaxies in LSST. This corresponds to a \dAtwoapproxCMB~smaller uncertainty on $A_2$ than currently available CMB constraints on anisotropic non-Gaussianity. We also show that constraints from Euclid will attain a similar precision.

This work is organized as follows. In Section \ref{sec:theory}, we summarize the tidal alignment model and its relation to anisotropic non-Gaussianity during inflation. In Section \ref{sec:surveys}, we describe the future surveys we consider for forecasting constraints on anisotropic non-Gaussianity from alignments. Section \ref{sec:fisher} describes the forecasting method, followed by the results in Section \ref{sec:results}. In Section \ref{sec:discuss}, we discuss the assumptions of our work and we suggest directions for future improvement. We conclude in Section \ref{sec:conclusion}. Throughout, we assume the following Planck \cite{Ade:2015} fiducial flat $\Lambda$CDM cosmology: $\Omega_{\rm b}h^2=0.022$, $\Omega_{\rm CDM}h^2=0.12$, $h=0.67$, $\Omega_K=0$, $\mathcal{A}_s=2.2\times10^{-9}$, $n_s=0.9645$, $k_p=0.05$ Mpc$^{-1}$ and we define $\Omega_m=\Omega_b+\Omega_{\rm CDM}$. 

\section{Inflation \& intrinsic alignments}
\label{sec:theory}

\subsection{Gaussian and non-Gaussian alignments}

In a Gaussian universe and on linear scales, the two components of the intrinsic galaxy shape are related to the tidal field via \citep{Catelan01}
\begin{equation}
  (\gamma_+^I,\gamma_\times^I) = b_I \frac{\left(k_x^2-k_y^2,2k_xk_y\right)}{k^2}\delta,
  \label{eq:ishape}
\end{equation}
where $b_I$ is a luminosity- and redshift-dependent bias that links the response of the galaxy ellipticity to the underlying tidal field of the large-scale structure. The $+$ component represents alignments tangential or orthogonal with respect to the separation vector. The $\times$ component is $45\,\deg$ rotated with respect to $+$. We will ignore the $\times$ component in this analysis, as it is null for both Gaussian and non-Gaussian initial conditions on linear scales (SCD15).  Further, any stochasticity in the bias relation Eq.~(\ref{eq:ishape}) only contributes at nonlinear order (we of course include the lowest order stochastic contribution, namely shape noise).  

The model Eq.~(\ref{eq:ishape}) applies to elliptical galaxies pressure-supported by the random orbits of their stars. On the other hand, disc galaxies are expected to be subject to alignments through a different mechanism: torques from the tidal field acting on their angular momentum.  This implies that the linear alignment of the type Eq.~(\ref{eq:ishape}) is absent \cite{lee/pen:2000,Catelan01}, so that large-scale alignments of late-type galaxies are expected to be highly suppressed on linear scales. For the purpose of this work, we will focus on elliptical alignments alone. Moreover, we will assume in general that elliptical galaxies can be identified by selecting on their red color, as a consequence of evolved stellar populations and low star formation rates.  An important observational result is that outer or inner regions of a galaxy are more or less sensitive to the tidal field, resulting in different $b_I$ values depending on the ellipticity estimator adopted \citep{Singh16,Tenneti15,Chisari15HzAGN,Chisari16HzAGN}.  

The effect of anisotropic non-Gaussianity on the primordial potential perturbation during matter domination is to modify the tidal field acting on biased shape tracers through the presence of additional anisotropic long-scale modes. This is analogous to the effect of {\it isotropic} non-Gaussianity on galaxy clustering \cite{dalal/etal,MacDonald08,Slosar08,PBSpaper}, but acting on galaxy shapes. In SCD15, we showed that anisotropic non-Gaussianity generates a scale-dependent bias in the $+$ component of intrinsic galaxy shapes. In the presence of anisotropic non-Gaussianity, $b_I$ transforms to
\begin{equation}
  b_I \rightarrow b_I+3b^{\rm NG}_IA_2\mathcal{M}^{-1}(k,z),
  \label{eq:bIfull}
  \end{equation}
where $\mathcal{M}$ describes the relation between $\delta$ and primordial potential $\phi$,
\begin{equation}
  \mathcal{M}(k,z) = \frac{2}{3}\frac{k^2T(k)D(z)}{\Omega_mH_0^2},
  \end{equation}
$H_0$ is the current value of the Hubble constant, $T(k)$ is the matter transfer function at $z=0$\footnote{The matter power spectrum and transfer function for the adopted cosmology in this work are obtained with the publicly available software CAMB \cite{camb}.}, and $b^{\rm NG}_I$ is a bias that quantifies the response of a galaxy shape to an anisotropic initial power spectrum of small-scale density perturbations. In SCD15, we adopted a value of $b^{\rm NG}_I$ that scales with $b_I(z)$, specifically
\begin{equation}
b_I^{\rm NG} = \tilde{b}_I^{\rm NG} b_I(z) \frac{D(z)}{D(0)}
  \end{equation}
and we set $\tilde{b}_I^{\rm NG}=1$. We will keep explicit track of $\tilde{b}_I^{\rm NG} $ in the analysis to highlight that intrinsic alignments are, in reality, sensitive to the product \Atwo, rather than $A_2$ alone.

On the other hand, gravitational lensing by the large-scale structure also contributes to modifying the ellipticity of a galaxy (with a shear $\gamma^G$), stretching it tangentially around a foreground overdensity. With the adopted sign convention, gravitational lensing sources a positive $+$ ellipticity around an overdensity, while for $b_I < 0$ intrinsic alignments source a negative $+$ ellipticity around the same overdensity and $A_2>0$ enhances the alignment towards a more negative signal. The overall ellipticity is given by
\begin{equation}
  \gamma = \gamma^I+\gamma^G+\gamma^{\rm rnd}
\end{equation}
where $\gamma^{\rm rnd}$ is a stochastic component uncorrelated with large-scale perturbations (and in particular $\gamma^I,\,\gamma^G$). In this work, we assume that two estimators $\gamma^{(1)},\,\gamma^{(2)}$ are available for each galaxy image, which, given the same lensing effect, differ in their intrinsic alignment contribution only. On large scales, we can assume the shot noise to be small and in the presence of non-Gaussianity with $A_2 \neq 0$,
\begin{equation}
  \frac{\gamma^{(1)}}{\gamma^{(2)}} = \frac{\gamma^{I,(1)}+\gamma^G}{\gamma^{I,(2)}+\gamma^G}
\label{eq:gammaratio}
\end{equation}
Note that since we assume that both the Gaussian and non-Gaussian component of the alignment bias are simply proportional to $b_I$, the ratio between different shape tracers is scale dependent only {\it due to the presence of lensing}. This allows us to use two different alignment tracers to obtain improved constraints on $A_2$ compared to a single alignment tracer. On the other hand, different values of $\tilde{b}^{\rm NG}_I$ for two alignment tracers could allow constraints on $A_2$ from alignments alone. The intrinsic alignment signal can be isolated from lensing through galaxy position-galaxy shape correlations when spectroscopic redshift information is available, i.e., galaxies are aligned towards tracers of the tidal field at the same redshift. The multiple-tracer technique, applied to constraining $A_0$ from clustering alone, does not suffer from this problem, because there the non-Gaussian bias $b_{\rm NG}^n$ scales as $b_1^n-1$. In that case, the ratio of the clustering bias is directly sensitive to $A_0$.

\subsection{Angular power spectra}

Intrinsic alignments can be best constrained from the cross-correlation of galaxy positions and galaxy shapes. On linear scales, the angular cross-spectrum between positions and shapes is given by 
\begin{equation}
C_{n\gamma}(l) = \frac2\pi \sqrt{\frac{(l-2)!}{(l+2)!}} \int dk\,k^2\:P_m(k) \left[F_l^I(k) +F_l^G(k)\right] F_l^n(k)\,, \label{eq:ClgE2}
\end{equation}
where $P_m$ is the linear matter power spectrum today, the relevant lensing ($G$), alignment ($I$) and clustering ($n$) kernels are
\begin{widetext}
\begin{eqnarray}
F_l^{G}(k) &=& \frac{1}{2}\frac{(l+2)!}{(l-2)!}  \int_0^{\chi_{\rm max}} d\chi\, k^2\,D_\Phi(k,z(\chi))\,\frac{j_l(x)}{x^2} \chi\int_{\chi}^{\chi_{\rm max}} d\tilde{\chi}  H(\tilde\chi) \frac{dN_G}{dz}\frac{(\tilde{\chi}-\chi)}{\tilde{\chi}} \,,\\
F_l^I(k) &=&  \frac{(l+2)!}{(l-2)!} \int dz \frac{dN_{\rm red}}{dz} 
\left[ b_I  + 3 b^{\rm NG}_I A_2 \M^{-1}(k, z) \right] \frac{D(z)}{D(0)}
\left[\frac{j_l(x)}{x^2}\right]_{x = k \chi(z)} \,,\\
F_l^n(k) &=& \int dz \frac{dN_n}{dz} \frac{D(z)}{D(0)}
\left[ b^n_1  + \frac12 b^n_{\rm NG} A_0 \M^{-1}(k, z) \right] 
j_l(x)\Big|_{x=k \chi(z)}\,,
\label{eq:FlGIn}
\end{eqnarray}
\end{widetext}
$dN_G/dz$ is the redshift distribution of source redshifts, $dN_n/dz$ is the redshift distribution of clustering redshifts, $dN_{red}/dz$ is the redshift distribution of red (aligned) galaxies, and $D_\Phi$ is given by
\begin{equation}
D_\Phi(k,z) = \frac{3H_0^2\Omega_{m}}{k^2} \frac{(1+z) D(z)}{D(0)}\,.
\end{equation}
Here, $b_1^n$ is the clustering bias, and we assume the standard universal mass function prediction for the scale-dependent bias due to isotropic non-Gaussianity, $b^n_{\rm NG} = (b^n_1-1) \delta_c$ where $\delta_c=1.686$ is the critical density.  In order to break the degeneracy between $A_0$ and $A_2$, which yield similar contributions to $C_{n\gamma}(l)$, we also use the autocorrelation of galaxy number counts. The clustering angular power spectrum is given by
\begin{equation}
C_{nn}(l) = \frac2\pi \int dk\,k^2 \:P_m(k) |F_l^n(k)|^2\,.
\label{eq:Clgg}
\end{equation}

In addition, the shape-shape correlation, which includes both the effects of lensing and alignments, and their cross-correlations, is given by
\begin{eqnarray}
C_{\g\g}(l) &=& \frac{2}{\pi} \frac{(l-2)!}{(l+2)!} \int dk\,k^2\,P_m(k) |F_l^I(k)+F_l^{G}(k)|^2\,.
\label{eq:ClSS}
\end{eqnarray}

For all angular power spectra, we adopt the Limber approximation \cite{Limber} for multipoles larger than $l=50$.

\section{Future large-scale structure surveys}
\label{sec:surveys}

Several experiments will see first light in the next decade with the goal to survey large contiguous areas of the sky. Constraints on anisotropic non-Gaussianity from alignments will particularly benefit from large angular coverage, as the number of Fourier modes probed increases as $f_{\rm sky}$, where $f_{\rm sky}$ is the fractional sky coverage. The increased depth of this next generation of surveys will also greatly improve on the cosmological volume accessible to current surveys by probing higher redshifts. We consider here two such experiments: LSST (from the ground) and Euclid (in space). LSST has slightly increased depth and probes higher redshifts than Euclid; but the latter benefits from a more stable point spread function for measuring galaxy ellipticities, free from atmospheric smearing.

\subsection{Large Synoptic Survey Telescope}

LSST will constrain dark energy from gravitational lensing in the redshift range from $z=0.1$ to $z=3$ over $18,000$ sq. deg. The `gold sample' of LSST galaxies with signal-to-noise ratio $>20$ and measured shears will be used for this measurement \cite{Chang13}. This sample will have $\sim 26$ galaxies per arcmin$^2$ to an effective $i$-band limiting magnitude of $25.3$ (AB). We adopt the redshift distribution from \cite{Chang13} in their fiducial scenario (Table 2), which has a median redshift of $z_m=0.83$. We assume that the root-mean-square dispersion of the ellipticities (``shape noise'') will be approximately $\sigma_{SN}=0.26$ per component.

\subsection{Euclid}

For Euclid, we assume the same characteristics as adopted in \cite{Krause15}. The area of the sky covered is $15,000$ sq. deg. The limiting magnitude in the $r$-band is $24.5$, allowing for a redshift coverage of $z=0$ to $z=2.5$. The number of galaxies with shapes is $30$ per arcmin$^2$. The redshift distribution of the sources is $\propto z^2 \exp[-(z/z_0)^{3/2}]$ with a mean of $\langle z\rangle=0.8$. We assume the same shape noise as for LSST.

\subsection{Red fraction \& intrinsic alignments}

We consider red galaxies as a proxy for elliptical galaxies subject to alignments. To model the red fraction as a function of redshift and luminosity we follow the approach outlined in the appendix of \cite{Joachimi11}. Given current observational constraints on the luminosity function of red galaxies, we estimate the redshift distribution of red galaxies, $dN_{red}/dz$, and the red fraction as
\begin{equation}
  f_{red}(z) = \frac{dN_{red}/dz}{dN_G/dz}\,,
  \label{eq:fred}
\end{equation}
where $dN_G/dz$ is the redshift distribution of the galaxies with shapes in each lensing survey.

The intrinsic alignment bias [Eq.~(\ref{eq:ishape})] is parameterized following standard convention as
\begin{equation}
b_I  = -A_I(L_r)C_1\rho_{{\rm crit},0}\Omega_m \frac{D(0)}{D(z)}
\label{eq:bI}
  \end{equation}
where $D(z)$ is the growth function of the matter perturbations, $C_1\rho_{{\rm crit},0}=0.0134$ is adopted by convention to match low redshift results \citep{Brown02}, and $A_I$ is a luminosity-dependent amplitude fixed to current observations. 
For the luminosity-dependent amplitude of intrinsic alignments of Eq. (\ref{eq:bI}), we adopt
\begin{equation}
A_I(L_r) = 5.76 \left( \frac{ L_r }{L_0}\right)^{1.13},
\label{eq:AIJ11}
\end{equation}
where $L_r$ is the average $r$-band luminosity at a given redshift and $L_0$ is a pivot luminosity; this $A_I(L_r)$ was measured by \cite{Joachimi11} from Luminous Red Galaxies (LRGs) in the {\it Sloan Digital Sky Survey} (SDSS, \cite{Eisenstein01}). Other observational constraints have been obtained by Ref. \cite{Singh15}, and are comparable. This will be our first and fiducial alignment tracer.

We will also consider a second tracer, with a value of $A_I^{(2)}$ that is $40\%$ higher than the fiducial one. This is inspired by the observational results of \cite{Singh16}, who find $40\%$ difference in alignment measurements of SDSS galaxies using isophotal shapes \cite{SDSSDR7} and shapes optimized for lensing \cite{Hirata03,Reyes12}. The authors of \cite{Singh16} interpreted this enhancement as a more efficient orientation of the outer isophotes of galaxies towards other massive galaxies. This could be a consequence of an enhanced tidal interaction of the large-scale structure with less tightly bound stars in the outskirts of galaxies. Notice that we assume that this difference between alignment estimators also applies to the non-Gaussian term of the alignment bias, i.e., to $b_I^{\rm NG}$. The results are sensitive to this assumption, as will be discussed in Section \ref{sec:discuss}.

\begin{figure}
  \includegraphics[width=0.45\textwidth]{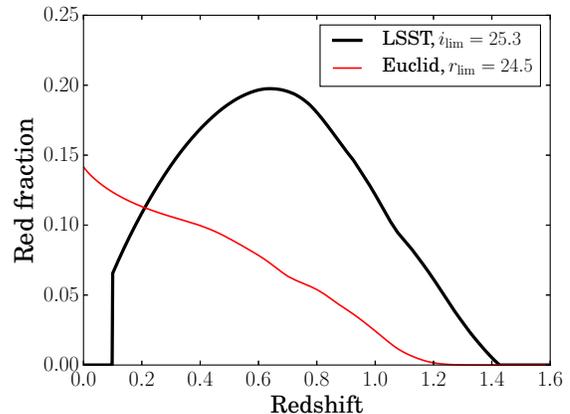}
  \caption{Fraction of red galaxies in LSST (black) and {\it Euclid} (red) as a function of redshift. The legend indicates the corresponding magnitude limits of the surveys.}
  \label{fig:redf}
\end{figure}

Following observational constraints from \cite{Mandelbaum11,Heymans13}, we will assume that disc (blue) galaxies do not exhibit detectable alignment on large scales. We model the fraction of blue galaxies as $f_{blue}(z) = 1- f_{red}(z)$. The red fraction, as a function of redshift is shown in Figure \ref{fig:redf}. Effectively, given the steepness of the red luminosity function with redshift \cite{Faber07}, at $z\gtrsim 1.4$, all galaxies are blue. For an LSST-like survey, we find that $\sim 2.6$ galaxies per sq. arcmin. are red and subject to alignments. For Euclid, this number drops to $\sim 1.6$ galaxies per sq. arcmin. Euclid is more shallow than LSST, which results in an increased fraction of red galaxies at low redshift. 

For red galaxies, we assume a fiducial clustering bias of $b_g^{red}=2$. For blue galaxies, we adopt an average clustering bias over the redshift range of the survey following the prescription of \cite{Alonso15}. In practice, the clustering bias is very close to $b_g^{blue}=2$ for both Euclid and LSST with this prescription.

\section{Fisher analysis}
\label{sec:fisher}

In SCD15, we used red galaxies from LSST to constrain \Atwo~using a combination of clustering ($n^{red}n^{red}$), galaxy-shape correlations ($n^{red}\gamma^{red}_{(1)}$) and shape-shape correlations ($\gamma^{red}_{(1)}\gamma^{red}_{(1)}$). Clustering and galaxy-shape correlations are sensitive to $A_0$ via the clustering bias of the galaxy tracers. Galaxy-shape and shape-shape correlations are sensitive to \Atwo~via intrinsic alignments. Gravitational lensing also contributes to galaxy-shape and shape-shape correlations and it is, in fact, dominant over alignments.

In this work, we extend the analysis performed in SCD15 to include the galaxy-shape and shape-shape correlations of a second intrinsic alignment tracer and between the two alignment tracers. We will assume that the gravitational lensing effect is the same for the two (that is, we assume that the shear responsivity for each shape estimator has been taken into account). We will also consider the clustering, galaxy-shape and shape-shape correlations of the blue galaxies in the survey. While these are insensitive to \Atwo, they provide constraints on the normalization of the matter power spectrum, $\sigma_8$, and on $A_0$. Finally, we will consider the impact of including the cross-correlation of blue galaxy positions with the red galaxy shapes for the two shape estimators. These cross-correlations provide additional constraints on both $A_0$ and \Atwo. The complete data vector of the 15 auto- and cross-correlations between 5 observables is 
\begin{eqnarray}
  {\bf D} = \bigg\{ C_{XY}(l)\,, & & \nonumber\\
  \mbox{with}\   XY &\in& \big\{ n^{red}n^{red},n^{red}\gamma^{red}_{(1)},\gamma^{red}_{(1)}\gamma^{red}_{(1)},\nonumber\\
    && n^{red}\gamma^{red}_{(2)},\gamma^{red}_{(1)}\gamma^{red}_{(2)},\gamma^{red}_{(2)}\gamma^{red}_{(2)},\nonumber\\
    && n^{blue}n^{blue},n^{blue}\gamma^{blue},\gamma^{blue}\gamma^{blue},\nonumber\\
    && n^{red}n^{blue},n^{blue}\gamma^{red}_{(1)},n^{blue}\gamma^{red}_{(2)},\nonumber\\
    && n^{red}\gamma^{blue},\gamma^{red}_{(1)}\gamma^{blue},\gamma^{red}_{(2)}\gamma^{blue}
\big\}
\bigg\}
\label{eq:Ddef}
\end{eqnarray}
where $n^{blue}$ refers to the positions of the blue galaxies, $\gamma^{red}_{(2)}$ is the second alignment tracer, and $\gamma^{blue}$ are the shapes of the blue galaxies (sensitive only to gravitational lensing).  Adopting a flat $\Lambda$CDM cosmology, the set of parameters to constrain using this data vector is ${\bf S}=\{A_0,\tilde{b}_I^{\rm NG}A_2,\sigma_8,b_{g}^{red},b_g^{blue},C_1\rho_{\rm crit,0},b_I^{r}C_1\rho_{\rm crit,0}\}$, where $b_I^r$ is the relative alignment bias of the two tracers. In addition we adopt a prior on $\sigma_8$ from Planck, i.e., $\sigma_8=0.831\pm0.013$ \cite[][Table 3, column 4]{Ade:2015}.  We do not include other cosmological parameters such as $\Omega_m,\,H_0$ in our Fisher forecast.  This is justified because we are interested in constraints on \Atwo~and $A_0$, which are determined by very large-scale correlations. On the other hand, $\Omega_m,\,H_0$ will be very well constrained from the shear two-point functions on small scales, in addition to independent probes such as baryon acoustic oscillations. 

Position and shape auto-correlations of the same tracer are subject to shot noise, while correlations of positions and shapes of the same tracer could be subject to correlated noise. For example, for blue galaxies, the correlated noise is given by $N_{n^{blue}\gamma^{blue}}=r_n\sigma_{SN}/\bar{n}^{blue}$, where $\bar{n}^{blue}$ is the surface density of blue galaxies per steradian. This correlated noise term is expected to be very small because the shape noise is dominated by the intrinsic dispersion in the ellipticities of galaxies, and is thus independent at first order of the fluctuations in galaxy number counts across the survey. Changing the value of $r_n$ does not have a significant impact on the results. We will consider the noise to be uncorrelated ($r_n=0$); and analogously for red galaxies.
On the other hand, the $\gamma^{red}_{(1)}$ and $\gamma^{red}_{(2)}$ cross-correlation is also subject to noise, as it results from two different shape estimators being applied to the same galaxy. We define the correlation coefficient of the noise between the shape tracers, \rng, such that $N_{\gamma^{red}_{(1)}\gamma^{red}_{(2)}}=r_n^{(\gamma)}\sigma_{SN}^{2}/\bar{n}^{red}$. The value of \rng, ranging from $0$ to $1$, can have significant impact on the cosmological constraints derived in this work.  Note that correlated noise can in fact be beneficial: for two random variates $A$ and $B$, the ratio $A/B$ [cf. Eq.~(\ref{eq:gammaratio})] has a smaller variance if they are positively correlated than in the uncorrelated case.

The covariance between two angular power spectra is given by 
\begin{equation}
  {\rm Cov}[C_{\alpha\beta}(l),C_{\gamma\delta}(l)]  = \frac{C_{\alpha\gamma}(l)C_{\beta\delta}(l)+C_{\alpha\delta}(l)C_{\beta\gamma}(l)}{{(2l+1)f_{\rm sky}}}
\end{equation}
where the indices $\{\alpha,\beta,\gamma,\delta\}$ run over number densities and shapes of the blue and red galaxies, including the first and second intrinsic alignment tracer. The contribution of the noise is taken into account in the angular power spectra.

The cosmological information is described by the Fisher matrix \cite{Tegmark97}, given by 
\begin{equation}
F_{\mu\nu} = \sum_l \frac{\partial {\bf D}(l)}{\partial p_{\mu}} {\rm Cov}^{-1}(l)  \frac{\partial {\bf D}(l)}{\partial p_{\nu}}
\end{equation}
where ${\rm Cov}$ is the covariance matrix and $\partial p$ indicates the partial derivative with respect to the parameters of interest, $\mu$ and $\nu$. These derivatives are obtained analytically from the expressions for $C_{\alpha\beta}(l)$, but the integration from Fourier to harmonic space is carried out numerically. The $1\sigma$ uncertainty in each parameter after marginalizing over all other parameters, are given by $\sigma_{\mu,{\rm full-margin}}=\sqrt{(F^{-1})_{\mu\mu}}$. Forecasted errors after marginalizing only over the bias parameters and $\sigma_8$ allow us to determine whether degeneracies between $A_0$ and \Atwo~exist. If $A_0$ and \Atwo~correspond to coordinates $1$ and $2$ of the Fisher matrix, the partially-marginalized bounds result from the following sub-matrix,
\begin{equation}
(G^{-1})_{i,j}=[F^{-1}]_{i=[1,2],j=[1,2]}
\end{equation}
and then the partially-marginalized uncertainties are
\begin{equation}
  \sigma_{i,{\rm part-margin}}=(G_{ii})^{-1/2}.
\end{equation}
    
\begin{figure}
\includegraphics[width=0.43\textwidth]{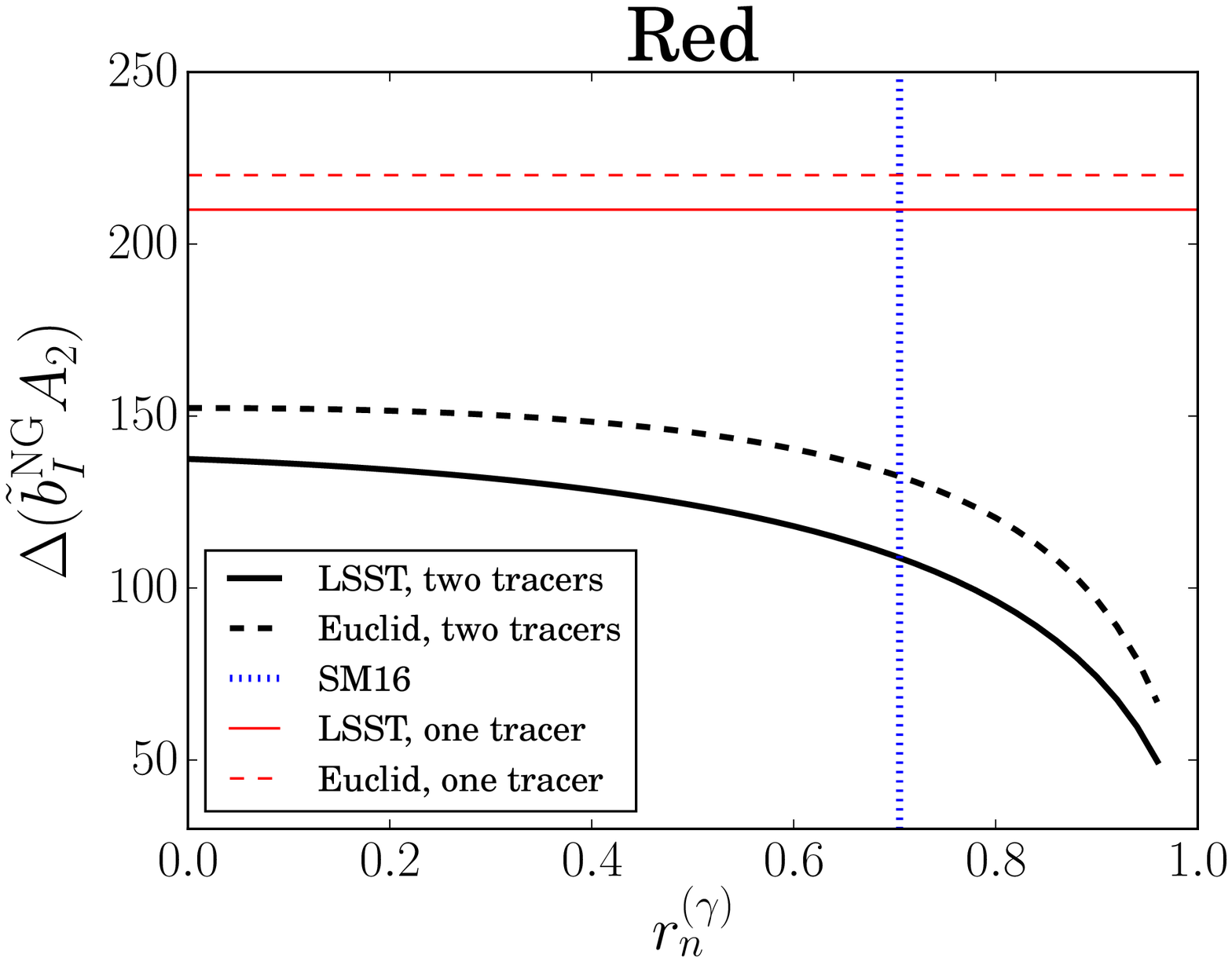} 
\includegraphics[width=0.43\textwidth]{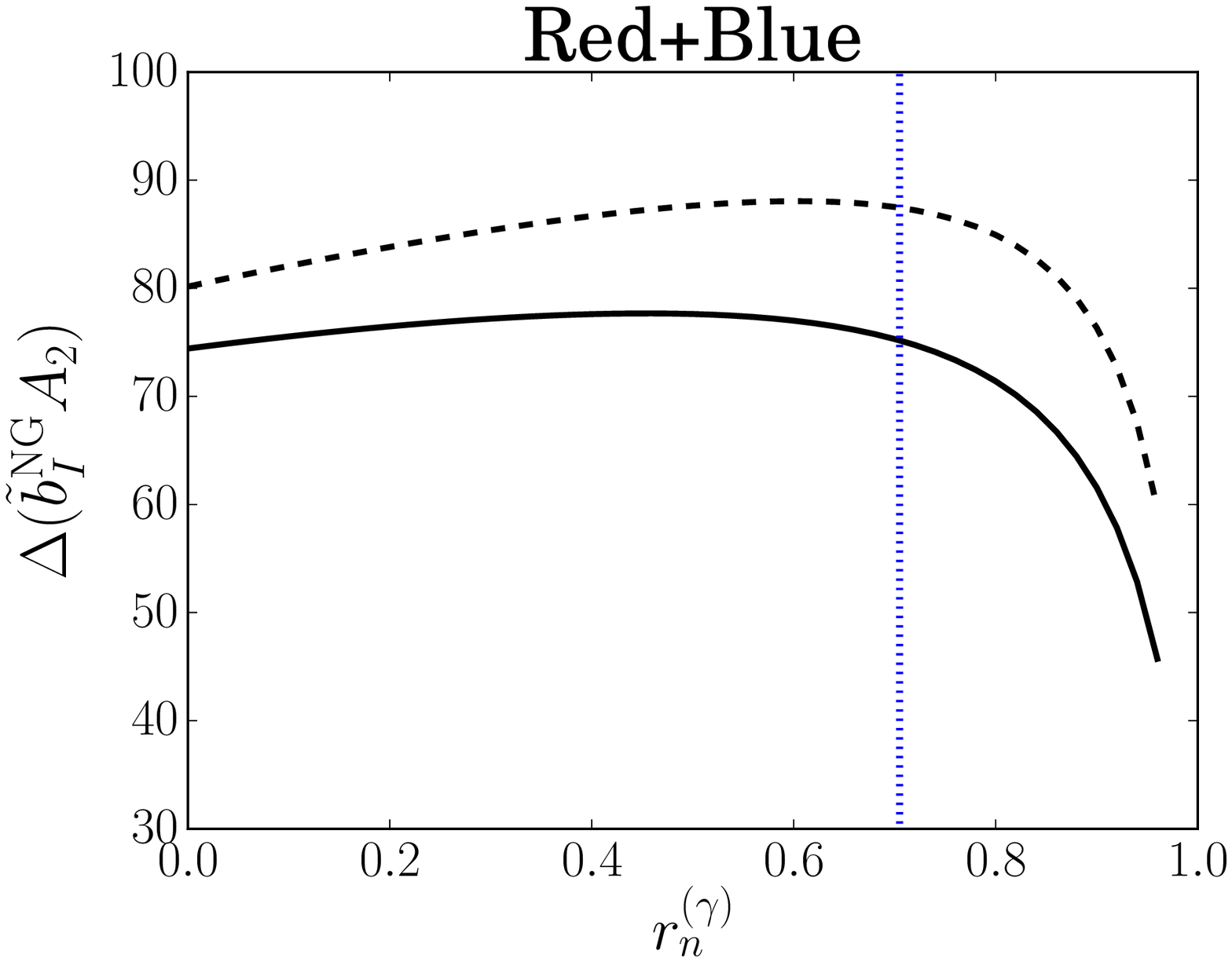} 
\includegraphics[width=0.43\textwidth]{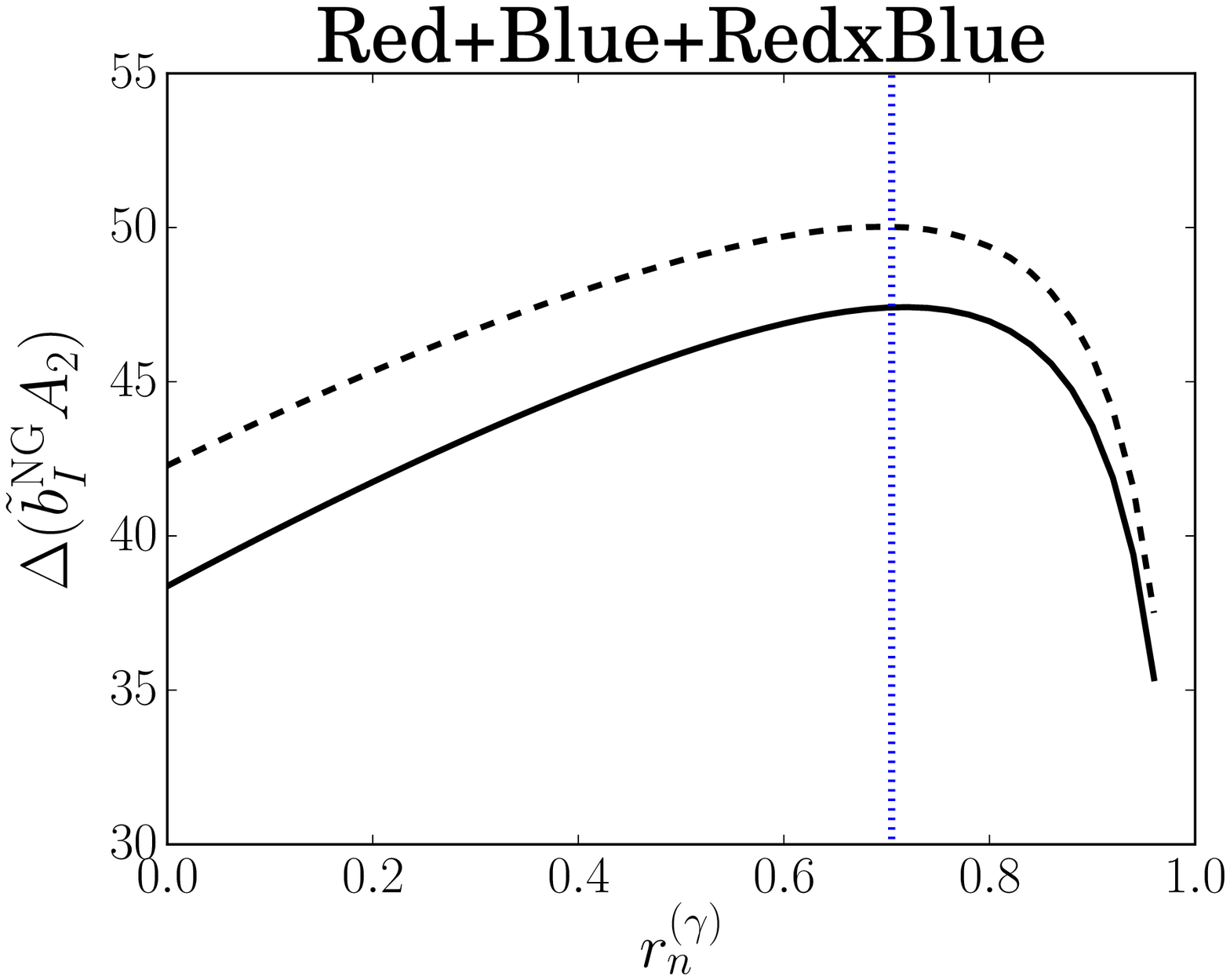}
\caption{Forecasted constraints on $A_2$ from LSST (solid) and {\it Euclid} (dashed) as a function of the correlation coefficient of the noise of the first and second alignment tracer, \rng. The top panel represents constraints from a single red tracer (red), and the new results with the addition of a second alignment tracer (black). The middle panel includes blue galaxy correlations. The bottom panel includes the blue position-red shape correlations. The blue dashed line labeled ``SM16'' refers to the value of \rng determined observationally by \cite{Singh16}. }
\label{fig:rnsA2}
\end{figure}
\begin{figure}
\includegraphics[width=0.45\textwidth]{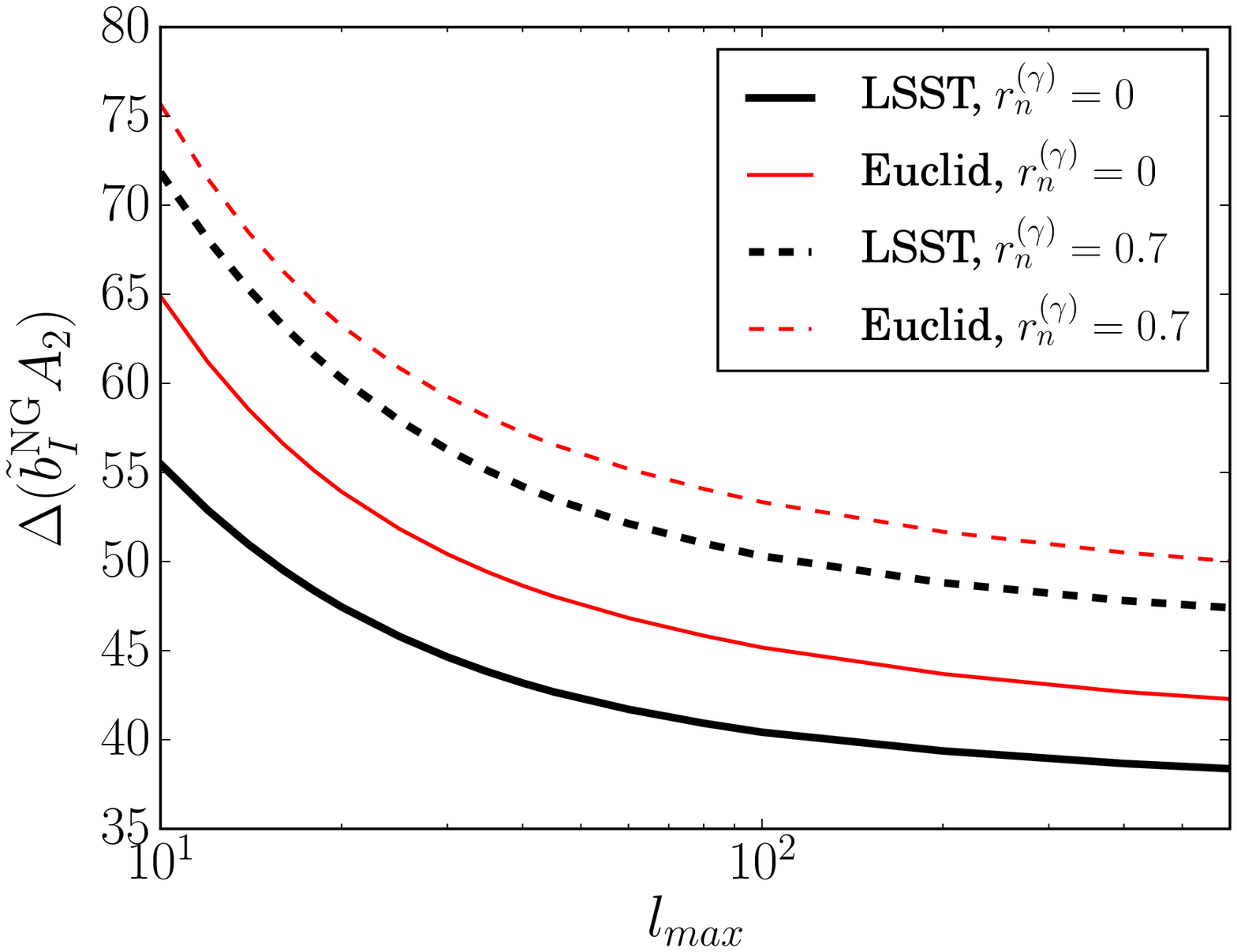} 
\includegraphics[width=0.45\textwidth]{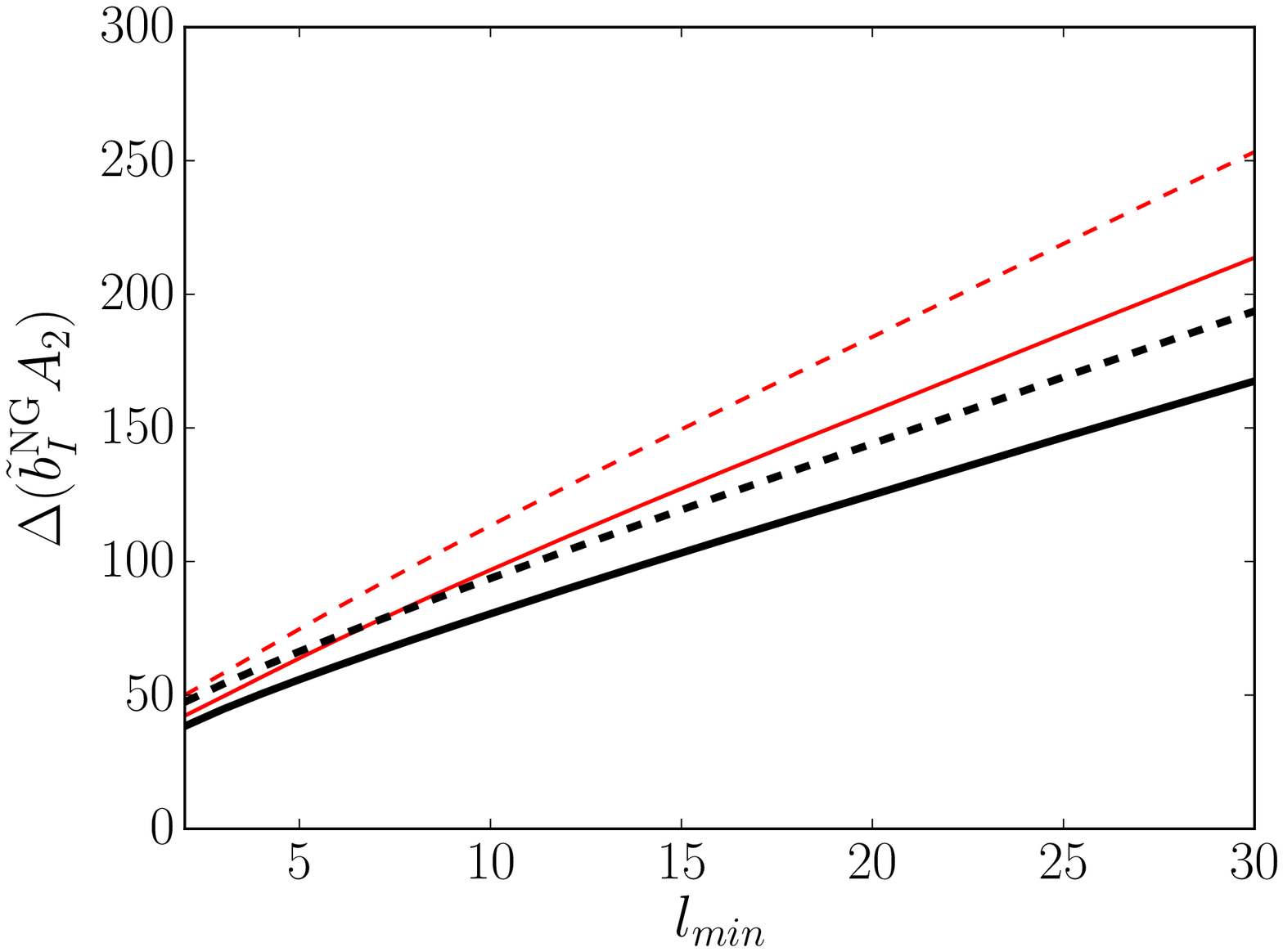} 
\caption{Forecasted uncertainty in \Atwo~from LSST as a function of maximum (top) or minimum (bottom) multipole probed. All cases include blue galaxies and their cross-correlations with the red sample. For the top panel, we fix the minimum multipole probed at $l=2$. Constraints improve with higher multipole, but the improvement as we approach nonlinear scales becomes marginal. For the bottom panel, we fix the maximum multipole probed to $l_{\rm max}=600$. The constraining power increases for smaller $l_{\rm min}$.}
\label{fig:lmax}
\end{figure}
\begin{figure*}
\includegraphics[width=0.47\textwidth]{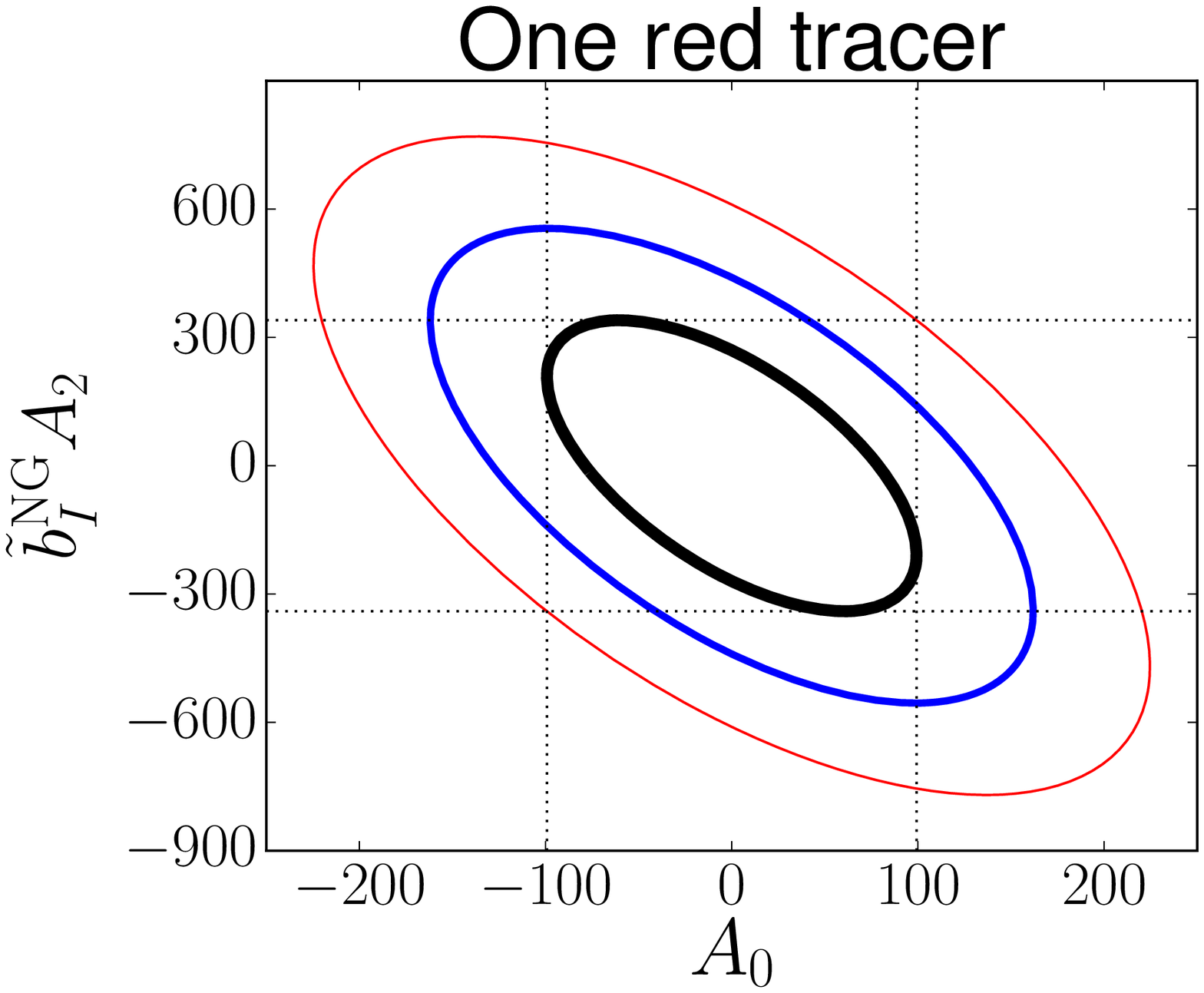}
\includegraphics[width=0.47\textwidth]{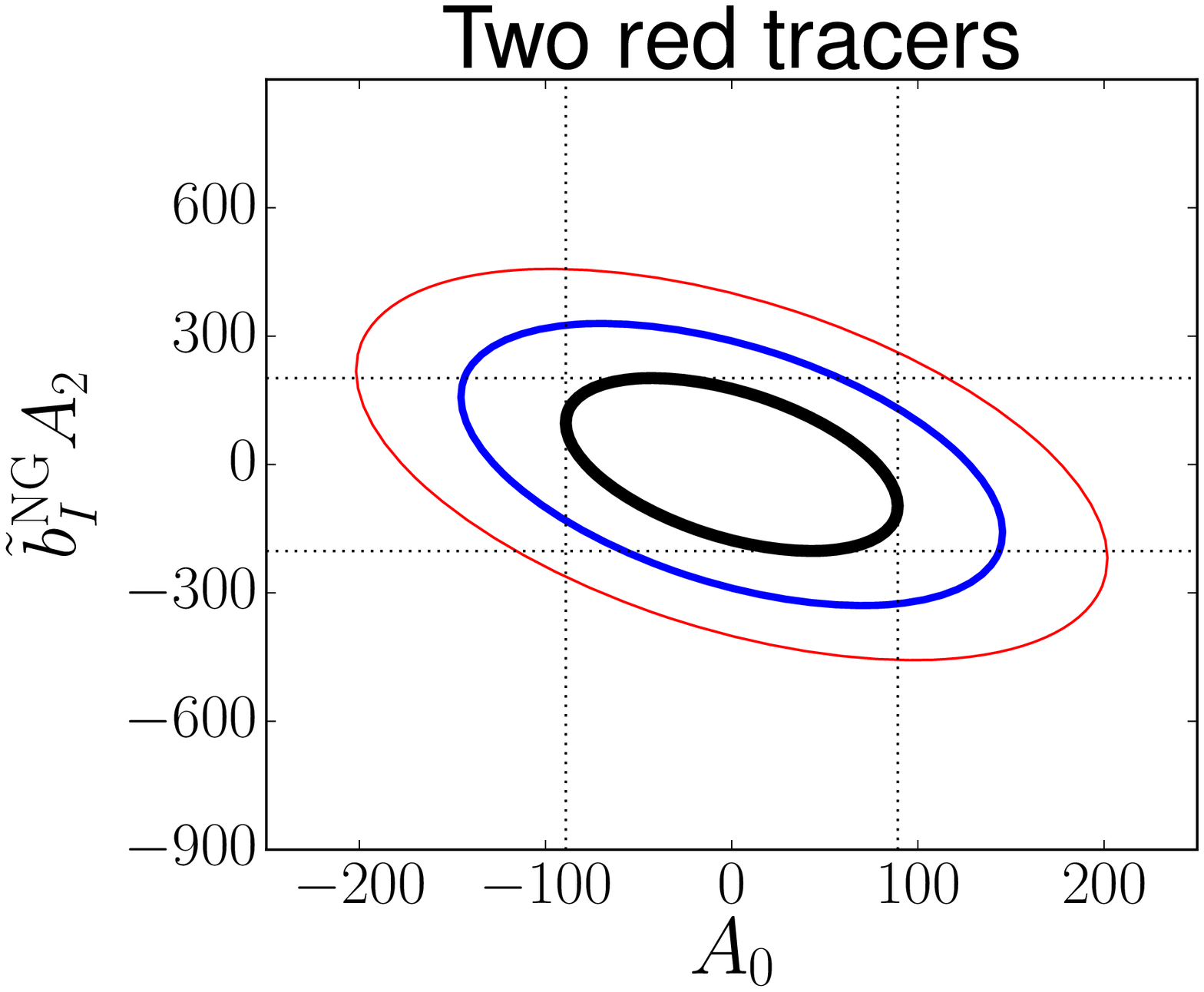}

\includegraphics[width=0.47\textwidth]{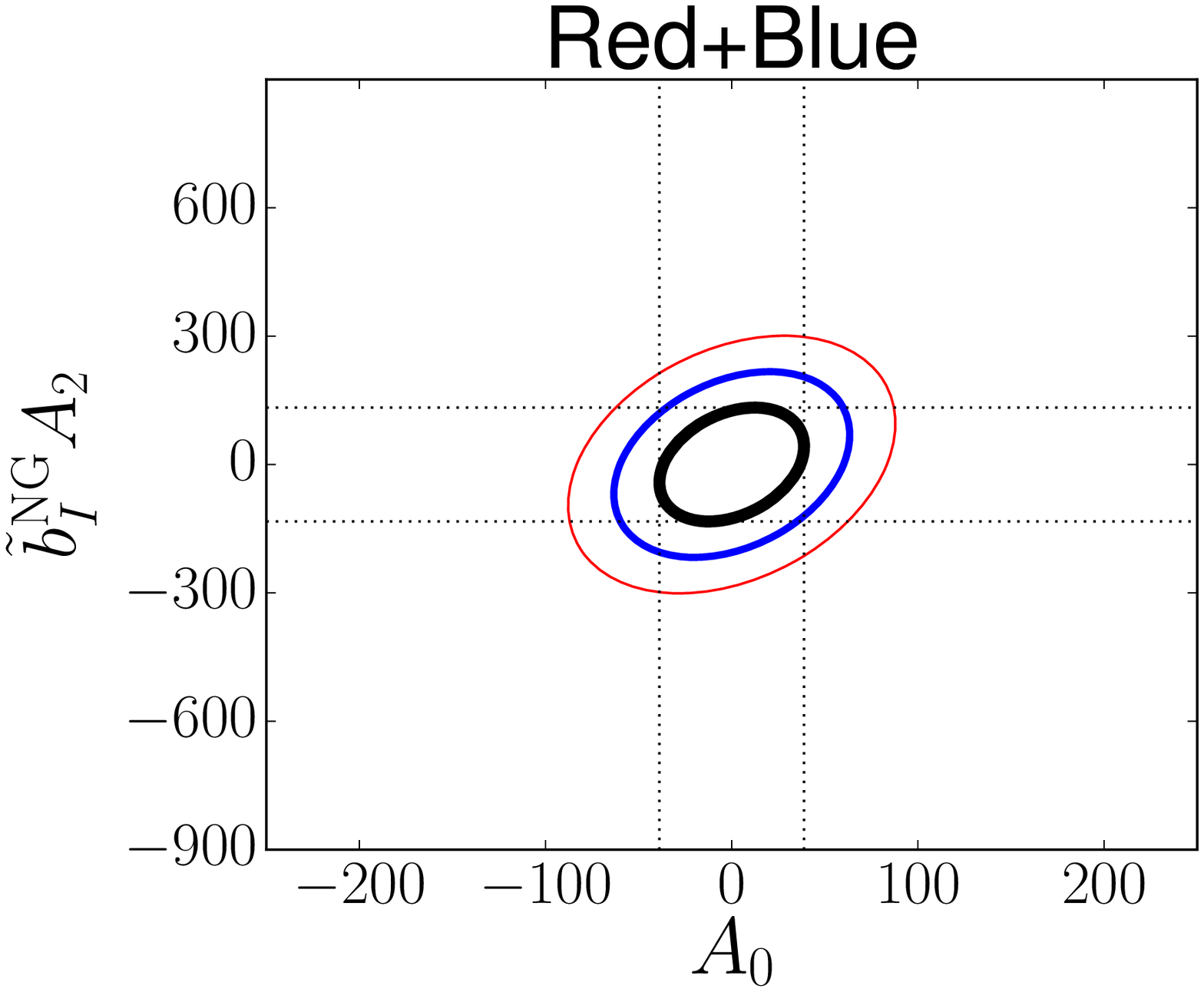}
\includegraphics[width=0.47\textwidth]{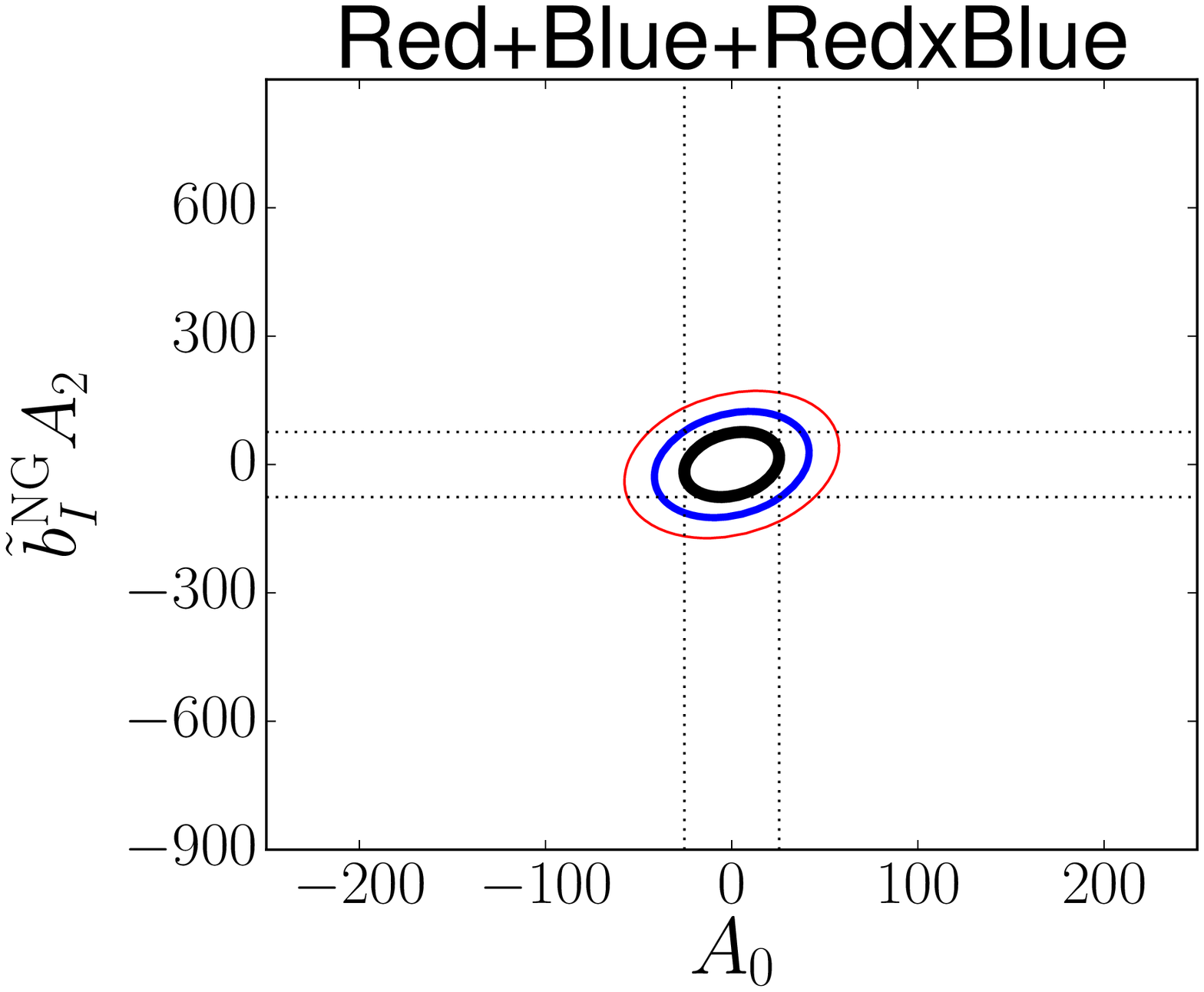}
\caption{Forecasted constraints on the $(A_0,\tilde{b}_I^{\rm NG}A_2)$ plane for Euclid. The top left panel represents the constraints from a single red tracer. More tracers are incorporated to the right: a second shape tracer (top right panel), blue galaxies (bottom left panel) and cross-correlations of red and blue galaxies (bottom right panel). The black, blue and red ellipses represent the ${1,2,3}\sigma$ contours, respectively. In these figures, we assume a level of correlated noise consistent with \cite{Singh16}.}
\label{fig:ellEuclid}
\end{figure*}

\section{Results}
\label{sec:results}

Starting from red galaxies as the only clustering and shape tracer, we find that the potential constraint on $A_2$ from LSST over $2<l<600$ is \dAtwo\AtwoLSSTred. The corresponding constraint on $A_0$ in this case is \dAzero\AzeroLSSTred. For Euclid, a single-tracer approach using only red galaxies yields: \dAtwo\AtwoEuclidred~and \dAzero\AzeroEuclidred. The constraints on both parameters are slightly worse than for LSST due to the smaller number density of red galaxies and the smaller cosmological volume at low redshift accessible through this tracer for Euclid.

Adding tracers progressively, we consider the following cases: the inclusion of a second shape tracer with $b_I^r=1.4$; the inclusion of auto-correlations of blue galaxies; the inclusion of cross-correlations of blue galaxy positions and red galaxy shapes. In each case, we vary \rng to assess the impact of correlated noise. The results are shown in Figure \ref{fig:rnsA2} for both surveys, where tracers are added from top to bottom. The vertical dashed line indicates the correlation coefficient of the noise \rng estimated from the measurements of \cite{Singh16}.
The top panel of Figure \ref{fig:rnsA2} shows that correlated noise improves the constraints on \Atwo~when two shape tracers subject to different alignment bias are used. This is in line with the results obtained by \cite{McDonald09}.  When including blue galaxies, the situation becomes more complicated, as they also contribute to cosmic variance cancelation.  This contribution apparently is more efficient when the noise of the two alignment tracers is uncorrelated.  For significantly correlated noise however, the constraints always improve for increasing $r_n^{(\gamma)}$.  

We also show how sensitive the constraint on \Atwo~is as a function of maximum multipole included in the Fisher calculation in the top panel of Figure \ref{fig:lmax}. Most of the information on non-Gaussianity comes from the largest scales, as expected from the $\propto k^{-2}$ dependence of the alignment bias [Eq.~(\ref{eq:bIfull})]. This simplifies the modelling required to extract the non-Gaussian signal, as a linear treatment of intrinsic alignments is sufficient at these scales. On the other hand, the details of the survey mask and Galactic foregrounds can complicate the extraction of the signal. Restricting to a range $10<l<600$, the potential constraints from LSST and Euclid in the case of multiple tracers are \dAtwo\AtwoLSSTrblten and \dAtwo\AtwoEuclidrblten, respectively, when the correlated noise is estimated from \cite{Singh16}. The dependence on the choice of the minimum multipole probed is shown in the bottom panel of Figure \ref{fig:lmax}.

The parameters $A_0$ and \Atwo~are degenerate through the position-intrinsic shape correlation, which is, on the other hand, the observable that provides the tightest constraints on \Atwo~by itself. A positive $A_0$ produces an enhancement of the clustering of red galaxies. Overall, this produces an enhancement of the position-shape correlation, as this observable is dominated by the contribution of lensing, rather than alignments, and these have opposite signs. On the other hand, a positive \Atwo~enhances the alignments, reducing the galaxy-lensing power spectrum. This effect is degenerate with a negative $A_0$. A second shape tracer is of help breaking the degeneracy between $A_0$ and \Atwo, as the effect of the latter is now more easily isolated. This progression can be seen in the top panels of Figure \ref{fig:ellEuclid} for the Euclid survey.

Once blue galaxies are incorporated into the data vector, a second tracer of the lensing power spectrum is available. This acts to effectively isolate the position-intrinsic shape correlation of red galaxies. As a result, $A_0$ and \Atwo~become degenerate in the other direction, as can be seen in the bottom left panel of Figure \ref{fig:ellEuclid}. A positive $A_0$ enhances the clustering bias of the red tracers, while a positive \Atwo~enhances their alignment. The constraints improve overall, with little change in the degeneracy as red and blue galaxy cross-correlations are added to the data vector. In this ultimate case, red and blue positions are sensitive to $A_0$, while cross-correlations involving red shapes improve \Atwo. This is shown in the bottom right panel of Figure \ref{fig:ellEuclid}.  Note that the relative impact of the various cross-correlations on the final constraints on $A_0,\,A_2$ does depend on the fiducial bias parameters for number counts and shapes.

Clearly, it would be interesting to investigate the use of the multi-tracer approach to mitigate the impact of intrinsic alignments on other cosmological parameters. However, by definition, sample variance cancelation only applies to scale-dependent signatures so that we do not expect significant direct improvements in $\sigma_8$ or the dark energy equation of state, for example.

\section{Discussion}
\label{sec:discuss}

We have explored the potential of a multi-tracer method for constraining anisotropic non-Gaussianity with the large-scale structure. This is the first application to intrinsic alignments of a method suggested originally by \cite{Seljak09} for galaxy clustering to improve constraints on {\it isotropic} non-Gaussianity, typically parametrized by the $f_{\rm NL}$ parameter ($\propto A_0$). Here, we have shown that the combination of two different shape tracers for intrinsic alignments greatly improves the constraint on the {\it anisotropic} non-Gaussianity parameter, \Atwo, from future galaxy surveys.

We expect the constraints on \Atwo and $A_0$ to improve further if a tomographic approach is adopted. Modes along the line-of-sight give additional information on $A_0$, reducing the constraints significantly \cite{Alonso15}. Moreover, tomography helps in distinguishing intrinsic alignments from weak lensing, as the alignments dominate the position-shape correlation for galaxies in the same redshift bin. 

Similarly to SCD15, we expect the constraints on \Atwo~to be sensitive to the evolution of the fraction of red galaxies with redshift and selection effects on the galaxies with shapes. In this work, the red fraction was determined based on current observational constraints on the luminosity function of red galaxies and the expected redshift distribution of galaxies in the optical survey. This estimate assumes that the red fraction is not significantly affected by selection effects in LSST or Euclid (image simulations as in \cite{Chang13} would be required to improve on this assumption). The current red galaxy luminosity function constraints only reach up to $z\sim 1$, much below the need of future surveys, and have significant uncertainty. Based on this data and extrapolating to higher redshifts, we find that red sources are unavailable beyond $z\sim 1.4$. If this assumption does not hold, a population of aligned galaxies at higher redshift could improve the constraints on \Atwo. Cosmological forecasts with intrinsic alignments should be revisited as additional observational constraints and image simulations become available. 

Our adopted intrinsic alignment bias [Eq.~(\ref{eq:AIJ11})] and the relative bias between the two alignment tracers rely on low redshift observations \cite{Joachimi11,Singh15,Singh16}. Hydrodynamic cosmological simulations suggest the persistence of a red aligned population at redshift as high as $z=3$ \cite{Chisari16HzAGN}, and a potential alignment signal for disc (blue) galaxies that increases to high redshift. If observations confirm these trends, it would be interesting to explore the impact of \Atwo~on blue galaxy alignments \cite{Catelan01,Hirata04,Schaefer15,Codis15,Larsen15} and the extension of the redshift coverage for red galaxies. This could have a significant impact on \Atwo~constraints. Ongoing galaxy surveys, like the Dark Energy Survey\footnote{\url{http://www.darkenergysurvey.org}}, the Kilo-Degree Survey\footnote{\url{http://kids.strw.leidenuniv.nl}} and HyperSuprime-Cam\footnote{\url{http://www.naoj.org/Projects/HSC/index.html}} are in a good position to start testing some of these hypotheses. 

The model applied in this work relies on a fully linear treatment for the tidal alignment. On small scales, nonlinear contributions are observed \cite{Joachimi11,Singh15} and a perturbation theory approach could be used to model them \cite{Blazek/etal:2015}. This potential extension to the modeling is outside of the scope of this work, but will be required to improve the constraints on cosmological parameters by accessing nonlinear scales in the future. Likewise, we have adopted a simple linear bias model for galaxy clustering, but scale-dependence is expected on small scales. 

We have also investigated the impact of an external prior on $A_0$, for instance from the Euclid galaxy redshift survey.  However, even an aggressive prior of $\sigma(A_0) = 4$, corresponding to $\sigma(f_{\rm NL})=1$, only leads to a marginal improvement on the constraint on $A_2$ when all cross-correlations are included.  This is because the degeneracy between $A_0$ and $A_2$ in this case is actually quite weak (lower right panel in Figure~\ref{fig:ellEuclid}).

Constraints on \Atwo~depend on the sensitivity of intrinsic shapes to an anisotropic power spectrum of initial density perturbations. While we have assumed that this response is similar in amplitude to the Gaussian alignment mechanism, the impact of adding the second red tracer varies depending on this assumption. If both alignment tracers responded similarly to a primordial anisotropic power spectrum, the second tracer in fact would not contribute to enhance the constraints on \Atwo. However, our multi-tracer approach would still provide a significant improvement on \Atwo~constraints compared to the single tracer method, due to the contribution of blue galaxies. These add information on $A_0$, which is degenerate with \Atwo. Hydrodynamic simulations of galaxy formation in the presence of an anisotropic initial power spectrum would be needed to investigate the response of different regions of a galaxy, i.e., different alignment estimators, to anisotropic non-Gaussianity in the spectrum of initial density perturbations.

In Figure \ref{fig:rnsA2} we showed that the constraints on \Atwo~are sensitive to $r_n^{(\gamma)}$, the correlation coefficient between the noise of the two shape tracers. For this work, we have adopted a value of $r_n^{(\gamma)}$ consistent with current intrinsic alignments observations. However, the value of $r_n^{(\gamma)}$ from future surveys could be explored by using image simulations that account for different shape measurement systematics, magnitude limit and varying galaxy populations.

\vskip 2pt
\section{Conclusion}
\label{sec:conclusion}

In the next decade, intrinsic alignments of galaxies could provide constraints on inflation complementary to the CMB bispectrum. Intrinsic shape estimators can be more sensitive to tidal alignments towards the outskirts of a galaxy, and different estimators can effectively be used in a multi-tracer technique in the spirit of \cite{Seljak09} to constrain anisotropic non-Gaussianity.

We have forecasted the impact of this method for two future weak lensing surveys, LSST and Euclid. Our results demonstrate that multi-traced intrinsic alignments, combined with lensing and clustering of blue galaxies, can yield constraints on the anisotropic non-Gaussianity parameter as low as \dAtwo\AtwoLSSTredxblue, corresponding to \dAtwoLSSTredxblue of the single tracer constraint. However, the impact of the atmospheric point-spread function on LSST might make it difficult to obtain two different alignment tracers, or it could result in a dependence of the relative alignment bias on apparent magnitude. Euclid is in a better position, due to the absence of the atmosphere, to perform shape measurements at different galactic radii. On the other hand, a tomographic approach has the potential of further shrinking the uncertainties in the non-Gaussianity parameters, as demonstrated by Ref. \cite{Alonso15} for $A_0$.

\acknowledgments
NEC is supported by a Beecroft fellowship. We are grateful to Sukhdeep Singh and Rachel Mandelbaum for providing the correlation coefficient of the noise of the two shape measurements used in \cite{Singh16}. We thank David Alonso for suggestions that helped improve this work. FS acknowledges support from the Marie Curie Career Integration Grant  (FP7-PEOPLE-2013-CIG) ``FundPhysicsAndLSS.''

\bibliographystyle{unsrt}
\bibliography{paper}

\end{document}